\documentclass[conference]{IEEEtran}
\usepackage{multicol}
\usepackage{lipsum,graphicx}
\usepackage{float}
\usepackage{epstopdf}
\usepackage{ifthen}
\usepackage[latin1]{inputenc}
\usepackage{amssymb}
\usepackage[cmex10]{amsmath}
\usepackage{dcolumn}
\usepackage[nolist]{acronym}

\DeclareGraphicsExtensions{.pdf,.jpeg,.png,.eps}
\usepackage{booktabs}

\usepackage[caption=false,font=footnotesize]{subfig}
\usepackage{comment}
\usepackage{color}
\usepackage{cite}

\usepackage{setspace}
\usepackage{mathrsfs}          
\usepackage{amsmath, amssymb, amsthm}
\usepackage{amsfonts}
\usepackage{mathrsfs}
\usepackage{multirow}        
\usepackage{framed}
\usepackage{psfrag}
\usepackage{graphicx}
\usepackage{textcomp, gensymb} 
\usepackage{algpseudocode} 

\newcolumntype{d}[1]{D{.}{\cdot}{#1} }

\usepackage{bm}
\usepackage{mathtools}
\usepackage{siunitx}

\newcommand{\PreserveBackslash}[1]{\let\temp=\\#1\let\\=\temp}
\newcolumntype{C}[1]{>{\PreserveBackslash\centering}p{#1}}
\newcolumntype{R}[1]{>{\PreserveBackslash\raggedleft}p{#1}}
\newcolumntype{L}[1]{>{\PreserveBackslash\raggedright}p{#1}}

\newcommand{\figref}[1]{Fig.~\ref{#1}}

\newcommand{\tabref}[1]{Table~\ref{#1}}

\usepackage{lipsum}
\usepackage{mathtools}
\usepackage{cuted}
\usepackage{stmaryrd}

\usepackage{tabularx}
\usepackage{tabulary}
\usepackage{tikz}
\usepackage{pgfplots}
\usepgfplotslibrary{units}
\pgfplotsset{compat=1.18}

\usepackage{algpseudocode}
\usepackage{algorithm}
\usepackage{hyperref}

\begin{document}
	\begin{acronym}
\acro{1G}{first generation}
\acro{2G}{second generation}
\acro{3G}{third generation}
\acro{3GPP}{Third Generation Partnership Project}
\acro{4G}{fourth generation}
\acro{5G}{fifth generation}
\acro{802.11}{IEEE 802.11 specifications}
\acro{A/D}{analog-to-digital}
\acro{ADC}{analog-to-digital}
\acro{AM}{amplitude modulation}
\acro{AP}{access point}
\acro{AR}{augmented reality}
\acro{ASIC}{application-specific integrated circuit}
\acro{ASIP}{Application Specific Integrated Processors}
\acro{AWGN}{additive white Gaussian noise}
\acro{BCJR}{Bahl, Cocke, Jelinek and Raviv}
\acro{BER}{bit error rate}
\acro{BFDM}{bi-orthogonal frequency division multiplexing}
\acro{BPSK}{binary phase shift keying}
\acro{BS}{base stations}
\acro{CA}{carrier aggregation}
\acro{CAF}{cyclic autocorrelation function}
\acro{Car-2-x}{car-to-car and car-to-infrastructure communication}
\acro{CAZAC}{constant amplitude zero autocorrelation waveform}
\acro{CB-FMT}{cyclic block filtered multitone}
\acro{CCDF}{complementary cumulative density function}
\acro{CDF}{cumulative density function}
\acro{CDMA}{code-division multiple access}
\acro{CFO}{carrier frequency offset}
\acro{CIR}{channel impulse response}
\acro{CM}{complex multiplication}
\acro{COFDM}{coded-\acs{OFDM}}
\acro{CoMP}{coordinated multi point}
\acro{COQAM}{cyclic OQAM}
\acro{CP}{cyclic prefix}
\acro{CR}{cognitive radio}
\acro{CRC}{cyclic redundancy check}
\acro{CRLB}{Cram\'{e}r-Rao lower bound}
\acro{CS}{cyclic suffix}
\acro{CSI}{channel state information}
\acro{CSMA}{carrier-sense multiple access}
\acro{CWCU}{component-wise conditionally unbiased}
\acro{D/A}{digital-to-analog}
\acro{D2D}{device-to-device}
\acro{DAC}{digital-to-analog}
\acro{DC}{direct current}
\acro{DFE}{decision feedback equalizer}
\acro{DFT}{discrete Fourier transform}
\acro{DL}{downlink}
\acro{DMT}{discrete multitone}
\acro{DNN}{deep neural network}
\acro{DSA}{dynamic spectrum access}
\acro{DSL}{digital subscriber line}
\acro{DSP}{digital signal processor}
\acro{DTFT}{discrete-time Fourier transform}
\acro{DUT}{device under test}
\acro{DVB}{digital video broadcasting}
\acro{DVB-T}{terrestrial digital video broadcasting}
\acro{DWMT}{discrete wavelet multi tone}
\acro{DZT}{discrete Zak transform}
\acro{E2E}{end-to-end}
\acro{eNodeB}{evolved node b base station}
\acro{E-SNR}{effective signal-to-noise ratio}
\acro{EVD}{eigenvalue decomposition}
\acro{FBMC}{filter bank multicarrier}
\acro{FD}{frequency-domain}
\acro{FDD}{frequency-division duplexing}
\acro{FDE}{frequency domain equalization}
\acro{FDM}{frequency division multiplex}
\acro{FDMA}{frequency-division multiple access}
\acro{FEC}{forward error correction}
\acro{FER}{frame error rate}
\acro{FFT}{fast Fourier transform}
\acro{FIR}{finite impulse response}
\acro{FM}		{frequency modulation}
\acro{FMT}{filtered multi tone}
\acro{FO}{frequency offset}
\acro{F-OFDM}{filtered-\acs{OFDM}}
\acro{FPGA}{field programmable gate array}
\acro{FSC}{frequency selective channel}
\acro{FS-OQAM-GFDM}{frequency-shift OQAM-GFDM}
\acro{FT}{Fourier transform}
\acro{FTD}{fractional time delay}
\acro{FTN}{faster-than-Nyquist signaling}
\acro{GFDM}{generalized frequency division multiplexing}
\acro{GFDMA}{generalized frequency division multiple access}
\acro{GMC-CDM}{generalized	multicarrier code-division multiplexing}
\acro{GNSS}{global navigation satellite system}
\acro{GPS}{global positioning system}
\acro{GPSDO}{GPS disciplined oscillator}
\acro{GS}{guard symbols}
\acro{GSM}{Groupe Sp\'{e}cial Mobile}
\acro{GUI}{graphical user interface}
\acro{H2H}{human-to-human}
\acro{H2M}{human-to-machine}
\acro{HTC}{human type communication}
\acro{I}{in-phase}
\acro{i.i.d.}{independent and identically distributed}
\acro{IB}{in-band}
\acro{IBI}{inter-block interference}
\acro{IC}{interference cancellation}
\acro{ICI}{inter-carrier interference}
\acro{ICT}{information and communication technologies}
\acro{ICV}{information coefficient vector}
\acro{IDFT}{inverse discrete Fourier transform}
\acro{IDMA}{interleave division multiple access}
\acro{IEEE}{institute of electrical and electronics engineers}
\acro{IF}{intermediate frequency}
\acro{IFFT}{inverse fast Fourier transform}
\acro{IoT}{Internet of Things}
\acro{IOTA}{isotropic orthogonal transform algorithm}
\acro{IP}{internet protocole}
\acro{IP-core}{intellectual property core}
\acro{ISDB-T}{terrestrial integrated services digital broadcasting}
\acro{ISDN}{integrated services digital network}
\acro{ISI}{inter-symbol interference}
\acro{ITU}{International Telecommunication Union}
\acro{IUI}{inter-user interference}
\acro{LAN}{local area netwrok}
\acro{LLR}{log-likelihood ratio}
\acro{LMMSE}{linear minimum mean square error}
\acro{LNA}{low noise amplifier}
\acro{LO}{local oscillator}
\acro{LOS}{line-of-sight}
\acro{LoS}{line of sight}
\acro{LP}{low-pass}
\acro{LPF}{low-pass filter}
\acro{LS}{least squares}
\acro{LTE}{long term evolution}
\acro{LTE-A}{LTE-Advanced}
\acro{LTIV}{linear time invariant}
\acro{LTV}{linear time variant}
\acro{LUT}{lookup table}
\acro{M2M}{machine-to-machine}
\acro{MA}{multiple access}
\acro{MAC}{multiple access control}
\acro{MAP}{maximum a posteriori}
\acro{MC}{multicarrier}
\acro{MCA}{multicarrier access}
\acro{MCM}{multicarrier modulation}
\acro{MCS}{modulation coding scheme}
\acro{MF}{matched filter}
\acro{MF-SIC}{matched filter with successive interference cancellation}
\acro{MIMO}{multiple-input multiple-output}
\acro{MISO}{multiple-input single-output}
\acro{ML}{machien learning}
\acro{MLD}{maximum likelihood detection}
\acro{MLE}{maximum likelihood estimator}
\acro{MMSE}{minimum mean squared error}
\acro{MRC}{maximum ratio combining}
\acro{MS}{mobile stations}
\acro{MSE}{mean squared error}
\acro{MSK}{Minimum-shift keying}
\acro{MSSS}[MSSS]	{mean-square signal separation}
\acro{MTC}{machine type communication}
\acro{MU}{multi user}
\acro{MVUE}{minimum variance unbiased estimator}
\acro{NEF}{noise enhancement factor}
\acro{NLOS}{non-line-of-sight}
\acro{NMSE}{normalized mean-squared error}
\acro{NOMA}{non-orthogonal multiple access}
\acro{NPR}{near-perfect reconstruction}
\acro{NRZ}{non-return-to-zero}
\acro{OFDM}{orthogonal frequency division multiplexing}
\acro{OFDMA}{orthogonal frequency division multiple access}
\acro{OOB}{out-of-band}
\acro{OQAM}{offset quadrature amplitude modulation}
\acro{OQPSK}{offset quadrature phase shift keying}
\acro{OTFS}{orthogonal time frequency space}
\acro{PA}{power amplifier}
\acro{PAM}{pulse amplitude modulation}
\acro{PAPR}{peak-to-average power ratio}
\acro{PC-CC}{parallel concatenated convolutional code}
\acro{PCP}{pseudo-circular pre/post-amble}
\acro{PD}{probability of detection}
\acro{pdf}{probability density function}
\acro{PDF}{probability distribution function}
\acro{PDP}{power delay profile}
\acro{PFA}{probability of false alarm}
\acro{PHY}{physical layer}
\acro{PIC}{parallel interference cancellation}
\acro{PLC}{power line communication}
\acro{PMF}{probability mass function}
\acro{PN}{pseudo noise}
\acro{ppm}{parts per million}
\acro{PPS}{pulse per second}
\acro{PRB}{physical resource block}
\acro{PRB}{physical resource block}
\acro{PSD}{power spectral density}
\acro{Q}{quadrature-phase}
\acro{QAM}{quadrature amplitude modulation}
\acro{QoS}{quality of service}
\acro{QPSK}{quadrature phase shift keying}
\acro{R/W}{read-or-write}
\acro{RAM}{random-access memmory}
\acro{RAN}{radio access network}
\acro{RAT}{radio access technologies}
\acro{RC}{raised cosine}
\acro{RF}{radio frequency}
\acro{rms}{root mean square}
\acro{RRC}{root raised cosine}
\acro{RW}{read-and-write}
\acro{SC}{single-carrier}
\acro{SCA}{single-carrier access}
\acro{SC-FDE}{single-carrier with frequency domain equalization}
\acro{SC-FDM}{single-carrier frequency division multiplexing}
\acro{SC-FDMA}{single-carrier frequency division multiple access}
\acro{SD}{sphere decoding}
\acro{SDD}{space-division duplexing}
\acro{SDMA}{space division multiple access}
\acro{SDR}{software-defined radio}
\acro{SDW}{software-defined waveform}
\acro{SEFDM}{spectrally efficient frequency division multiplexing}
\acro{SE-FDM}{spectrally efficient frequency division multiplexing}
\acro{SER}{symbol error rate}
\acro{SIC}{successive interference cancellation}
\acro{SINR}{signal-to-interference-plus-noise ratio}
\acro{SIR}{signal-to-interference ratio}
\acro{SISO}{single-input, single-output}
\acro{SMS}{Short Message Service}
\acro{SNR}{signal-to-noise ratio}
\acro{SSB}{single-sideband}
\acro{STC}{space-time coding}
\acro{STFT}{short-time Fourier transform}
\acro{STO}{symbol time offset}
\acro{SU}{single user}
\acro{SVD}{singular value decomposition}
\acro{TD}{time-domain}	
\acro{TDD}{time-division duplexing}
\acro{TDMA}{time-division multiple access}
\acro{TFL}{time-frequency localization}
\acro{TO}{time offset}
\acro{TS-OQAM-GFDM}{time-shifted OQAM-GFDM}
\acro{UE}{user equipment}
\acro{UFMC}{universally filtered multicarrier}
\acro{UL}{uplink}
\acro{US}{uncorrelated scattering}
\acro{USB}{universal serial bus}
\acro{UW}{unique word}
\acro{VLC}{visible light communications}
\acro{VR}{virtual reality}
\acro{WCP}{windowing and \acs{CP}}	
\acro{WHT}{Walsh-Hadamard transform}
\acro{WiMAX}{worldwide interoperability for microwave access}
\acro{WLAN}{wireless local area network}
\acro{W-OFDM}{windowed-\acs{OFDM}}	
\acro{WOLA}{windowing and overlapping}	
\acro{WSS}{wide-sense stationary}
\acro{ZCT}{Zadoff-Chu transform}
\acro{ZF}{zero-forcing}
\acro{ZMCSCG}{zero-mean circularly-symmetric complex Gaussian}
\acro{ZP}{zero-padding}
\acro{ZT}{zero-tail}
\acro{URLLC}{ultra-reliable low-latency communications}
\acro{PLL}{phase-locked loop}
\acro{USRP}{universal software radio peripheral}
\acro{TX}{transmission}
\acro{REF}{reference}
\acro{PFD}{phase frequency detector}
\acro{LF}{loop filter}
\acro{VCO}{voltage-controlled oscillator}
\acro{TIE}{time interval error}
\acro{ACF}{autocorrelation function}
\acro{OU}{Ornstein-Uhlenbeck}
\acro{CCF}{cross-correlation function}
\acro{WSS}{wide-sense stationary}
\acro{SDE}{stochastic differential equation}

\acro{HSI}{human system interface}
\acro{HMI}{human machine interface}
\acro{VR} {visual reality} 
\acro{AGV}{automated guided vehicles}
\acro{MEC}{multiaccess edge cloud}
\acro{TI} {tactile Internet}
\acro{IMT}{ international mobile telecommunications}
\acro{GN}{gateway node}
\acro{CN}{control node}
\acro{NC}{network controller}
\acro{SN}{sensor node}
\acro{AN}{actuator node}
\acro{HN}{haptic node}
\acro{TD}{tactile devices}
\acro{SE}{supporting engine}
\acro{AI}{artificial intelligence}
\acro{TSM}{tactile service manager}
\acro{TTI}{transmission time interval}
\acro{NR}{new radio}
\acro{SDN}{software defined networking}
\acro{NFV}{ network function virtualization}
\acro{CPS}{cyber-physical system}
\acro{TSN}{Time-Sensitive Networking}
\acro{FEC}{forward error correction}
\acro{STC}{space-time  coding}
\acro{HARQ}{hybrid automatic repeat request}
\acro{CoMP} {Coordinated multipoint}
\acro{HIS}{human system interface }
\acro{RU}{radio unit}
\acro{CU}{central unit}
\acro{AoD} {angle of departure}
\end{acronym}
	\title{Estimating PLL Phase Noise Parameters from Measurements for System-Level Modeling} 
	
	\author{
		\IEEEauthorblockN{
			Carl Collmann, Ahmad Nimr, Gerhard Fettweis
			}
			
		\IEEEauthorblockA{
		Vodafone Chair Mobile Communications Systems, Technische Universit\"{a}t Dresden, Germany\\ \small\texttt{\{carl.collmann, ahmad.nimr,  gerhard.fettweis\}@tu-dresden.de}\\
		}
		}
	\maketitle
	\IEEEpeerreviewmaketitle
	\begin{abstract}

In current MIMO mobile communication systems, phase noise can significantly impair performance.
To allow for compensation of these impairments, accurate phase noise modeling is necessary.
Numerical modeling of the phase noise process at a \ac{PLL} output is established in the literature and commonly represented by an \ac{OU} process.
The corresponding spectrum can be represented by a multi-pole/zero model.
This work presents  a \ac{LS} method for estimating the \ac{PLL} parameters such as oscillator constants or \ac{PLL} bandwidth from a measured phase noise spectrum.
The method is applied on the MAX2870 and MAX2871 \ac{PLL} chips and parameter estimates such as oscillator constants and \ac{PLL} bandwidths are provided.
The resulting parameter set enables both time- and frequency-domain numerical simulations.

\end{abstract}

\acresetall
\section{Introduction}\label{sec:introduction}
The upcoming sixth generation (6G) of mobile communication networks aims to improve performance in \ac{MIMO} systems, with a focus on data transmission and sensing, which are impacted by the presence of hardware impairments \cite{6gvision23}.
For example, in \ac{MIMO} systems the presence of phase noise can degrade beamforming gain as it hinders the coherent combination of signals \cite{Heath_mimo_18}.
Therefore, the modeling of phase noise for specific hardware is of interest to mitigate its effects.

Existing works on phase noise can be split into three main categories: 1) research that focuses on numerical modeling and theory of phase noise \cite{3gppTR,9771968_Piemontese,1637183_herzel,mehrotra1031966}; 2) investigation into the application of phase noise, specifically its effect on communication systems \cite{6693756_khanzadi} or sensing performance \cite{Coll202501}; and 3) literature that focuses on hardware aspects of \ac{PLL} development and subsequent measurements or experimentation \cite{11030065_ahmad,8845251_quadri,9916251_yang,tschapek9721401}. 
Works that bridge these categories are rare and often face practical limitations.
For example 3GPP \cite{3gppTR} presents phase noise \ac{PSD} measurements at $\SI{29.5}{}$, $\SI{45}{}$ and $\SI{70}{GHz}$ with a pole-zero model and corresponding parameters.
However, this work has key limitations: the model parameters (for example figure 6.1.10-2 with parameters in corresponding table 6.1.10-1) are provided without an estimation method, limiting reproducibility; the models may not be applicable to different frequencies or \ac{PLL} architectures; and the model does not provide a link to the time-domain phase noise process.

This paper addresses the gap by presenting a method to estimate the parameters of a simplified phase noise model from measurement data.
The corresponding time- and frequency domain models are reported in our work \cite{collmann2025practicalanalysisunderstandingphase}.
The feasibility of this modeling approach has been demonstrated in previous work \cite{Coll202501}, where it was used to investigate the effect of phase noise on angle estimates.
Phase noise \ac{PSD} measurements are conducted for MAX2870 \cite{MAX2870} and MAX2871 \cite{MAX2871} \ac{PLL} synthesizers and the method for estimating the phase noise parameters is demonstrated.
These parameter estimates in conjunction with the provided model enable generation of phase noise processes with identical statistical properties to the measured hardware.

The remainder of the paper is organized as follows:
Section \ref{sec:syst_model} introduces the simplified phase noise PSD model.
Next, the conducted measurements and data pre-processing are described in section \ref{sec:measurement}.
The phase noise parameter estimation procedure and results are provided in section \ref{sec:parameter}.
Section \ref{sec:conclusion} concludes the paper with key findings and future applications.



\section{PLL Phase Noise PSD Model}\label{sec:syst_model} 

The \ac{SSB} phase noise \ac{PSD} of a \ac{PLL} can be described by a pole-zero model \cite{3gppTR}.
In this work, we adopt the model derived in our work \cite{collmann2025practicalanalysisunderstandingphase}, which is based on an \ac{OU} process representation of the \ac{PLL} output \cite{mehrotra1031966}.
The \ac{PSD} is expressed in terms of the offset frequency $\Delta f = f - f_0$, with $f_0$ denoting the oscillator frequency and is given by
\begin{align}
    \mathcal{L}(\Delta f) = &-10 \log_\text{10} \left(\pi f_\text{c,REF} \right) \label{eq:pn_model_simple}\\
    &+ 10\log_\text{10} \left[ \frac{1 + \left(\frac{\Delta f}{\Delta f_\text{PLL}}\right)^3}{1 + \left(\frac{\Delta f}{f_\text{c,REF}}\right)^3} \frac{1 + \left(\frac{\Delta f}{\Delta f_\text{NF}}\right)^3}{1 + \left(\frac{\Delta f}{B_\text{PLL}}\right)^3} \right] \nonumber.
\end{align}
The parameter $f_\text{c,REF}$ refers to the $\SI{3}{dB}$ cut-off frequency of the reference oscillator at the \ac{PLL} input.
It is related to the reference oscillator constant $c_\text{REF}$ by $f_\text{c,REF} = \pi f_0^2 c_\text{REF}$.
The \ac{PLL} output spectrum is flat from $\Delta f_\text{PLL}$ to \ac{PLL} bandwidth $B_\text{PLL}$.
This behavior arises from modeling the \ac{PLL} output timing jitter as an \ac{OU} process \cite{mehrotra1031966}.
Previous research has shown that this modeling provides good agreement for SiGe oscillators \cite{11030065_ahmad}.
The parameter $\Delta f_\text{NF}$ denotes the frequency at the intersection point between \ac{VCO} model \eqref{eq:vco_model} and the noise floor. 
The \ac{PLL} output spectrum asymptotically approaches that of the reference oscillator at small offsets and a free-running \ac{VCO} at large frequency offsets.
Their respective models \cite{collmann2025practicalanalysisunderstandingphase} are (with $i \in \{_\text{REF},_\text{VCO}\}$ for reference oscillator and \ac{VCO})
\begin{align}
    \mathcal{L}_\text{i}(\Delta f) = &-10 \log_\text{10} \left(\pi f_\text{c,i}\right) -10 \log_\text{10} \left[ 1 + \left( \frac{\Delta f}{f_\text{c,i}}\right)^3 \right] \label{eq:vco_model}.
\end{align}


\section{Measurement Setup and Data Collection}\label{sec:measurement}

\begin{figure}[tb]
    \centering
    \includegraphics[width=\linewidth]{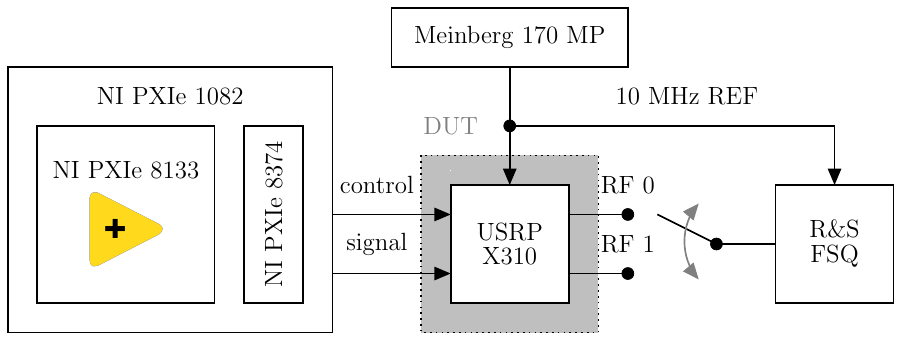}
    \caption{Setup for \ac{PSD} measurement of MAX2870/71 \ac{PLL} synthesizers integrated on UBX/CBX daughterboards of \ac{USRP} X310 \cite{collmann2025practicalanalysisunderstandingphase}}
    \label{fig:meas_setup}
\end{figure}

Measurements of the phase noise spectrum density are conducted for different models of the \ac{USRP} X310 listed in \cite{data_bpmg_pc85_24}. The \ac{USRP} models considered feature two different daughterboard modules, CBX and UBX, with MAX2870 \cite{MAX2870} and MAX2871 \cite{MAX2871} \ac{PLL}'s respectively, as frequency synthesizing circuits. 
The measurements are conducted with a R\&S FSQ8 spectrum analyzer and the \ac{DUT} locked to a $\SI{10}{MHz}$ reference signal provided by a Meinberg 170 MP GPS receiver. 
The measurements are obtained for the range $\Delta f \in \{\SI{100}{Hz},\SI{10}{MHz} \}$ with $1\%$ resolution bandwidth for each increment (e.g. $\SI{30}{Hz}$ for the $\SI{1}{kHz}$ to $\SI{3}{kHz}$ segment) with a total sweep duration of $\SI{24.7}{s}$.
Each measured \ac{PSD} is averaged over $10$ observations using the FSQ8's linear smoothing option.
The spectrum analyzer's noise floor was measured at approximately $\SI{-153}{dBc/Hz}$ for offset frequencies greater than $\SI{300}{kHz}$ by terminating the spectrum analyzer with a $\SI{50}{\Omega}$ load.
The \ac{DUT} phase noise remained at least $\SI{8}{dB}$ above this noise floor, ensuring measurement validity.
The USRP is configured to output a continuous wave at $\SI{2}{GHz}$ with an output power of $\SI{0}{dBm}$.
A schematic of the measurement setup is provided in \figref{fig:meas_setup} and the datasets are available at \cite{data_bpmg_pc85_24}.

\begin{figure}[tb]
    \centering
    \includegraphics[width=\linewidth]{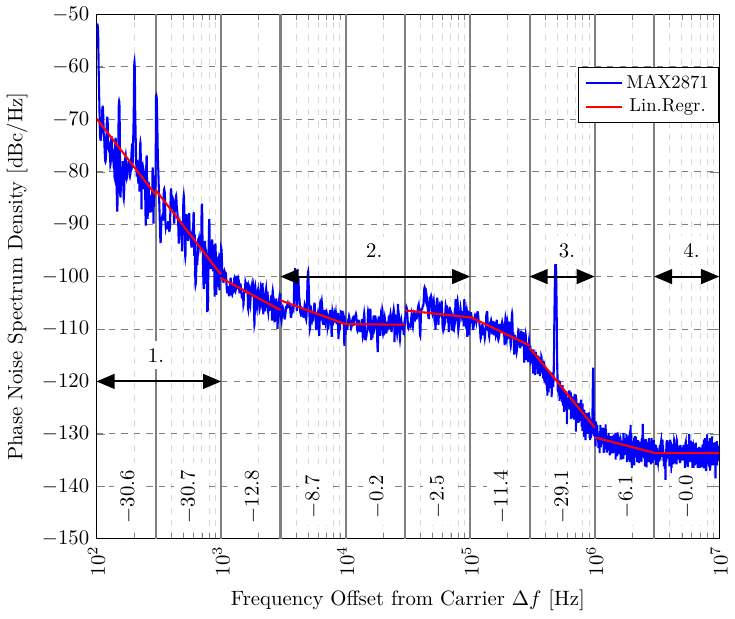}
    \caption{Measured phase noise spectrum for \ac{USRP} 2944R with UBX daughterboard at $f_\text{c}=\SI{2}{GHz}$ and piece-wise linear regression with slope $\hat{r}[1]$ \eqref{eq:lin_regr} at the bottom of the plot}
    \label{fig:pn_lin_regr}
\end{figure}

As illustrated in \figref{fig:pn_lin_regr}, the measured \ac{PSD} of the \ac{PLL} output is characterized by the following parts: 1) reference dominant; 2) \ac{PLL} in-band noise floor; 3) \ac{VCO} dominant; and 4) noise floor.
This process is shown for a measured \ac{PSD} of a \ac{USRP} 2944R (blue).
The characterization is performed by evaluating the slope of the spectrum for half a decade frequency segments.
This slope is calculated using a piece-wise linear regression (red) given by
\begin{align}
    \hat{\bm{r}} = \left( \textbf{X}^{T} \textbf{X} \right)^{-1} \textbf{X}^{T} \textbf{y}. 
    \label{eq:lin_regr}
\end{align}
The matrix
\begin{align}
    \textbf{X} = \begin{bmatrix}
        1 & \cdots & 1\\
        \log_{10} (f_{1}) & \cdots & \log_{10} (f_{N})
    \end{bmatrix}
\end{align}
contains the logarithmized bounds of the segment for which the regression is calculated, while $\textbf{y} = [\mathcal{L}(f_{1}), \cdots, \mathcal{L}(f_{N})]$ represents the corresponding magnitude values of the phase noise spectrum at $N$ discrete points inside a frequency segment.
The following sections are distinguished for a specific \ac{PLL} MAX2871 by grouping segments with similar slope:
\begin{itemize}
    \item[1.] $f < \SI{1}{kHz}$: \ac{PLL} follows reference oscillator
    \item[2.] $\SI{3}{kHz} \leq f \leq \SI{100}{kHz}$: \ac{PLL} in-band noise floor
    \item[3.] $\SI{300}{kHz} ~\leq f\leq ~\SI{1}{MHz}$: \ac{PLL} follows \ac{VCO}
    \item[4.] $f > \SI{3}{MHz}$: noise floor
\end{itemize}
The transition regions between the bands (for example the segment $\Delta f \in \{ \SI{1}{kHz}, \SI{3}{kHz}\}$) are excluded from the characterization to avoid fitting errors near the corner frequencies.

	\section{Parameter Estimation}\label{sec:parameter}
In this section, the measured \acp{PSD} are used to estimate the parameters of the simplified model \eqref{eq:pn_model_simple}.
The primary parameters of interest are the oscillator constants $c_\text{VCO}$ and $c_\text{REF}$, their corresponding $\SI{3}{dB}$ cut-off frequencies, and the \ac{PLL} loop bandwidth $B_\text{PLL}$.
Estimators for the phase noise model parameters are presented and explicitly calculated for the recorded data-set.
First, the procedure is demonstrated on a single USRP with the UBX daughterboard, featuring the MAX2871 \ac{PLL}.
Then, parameter estimates are presented for all devices in the data-set, covering both CBX and UBX  daughterboards.

\subsection{Estimators for Phase Noise Model Parameters}
\label{sec:param_parameter_est}

To estimate the oscillator constants, it is necessary to first estimate the cut-off frequencies for the \ac{LP} model that represents the phase noise characteristic of \ac{VCO} and the reference oscillator from \eqref{eq:vco_model}, which can be approximated as 
\begin{align}
    \mathcal{L}_\text{i}(\Delta f)
    & \approx 20 \log_\text{10} \left(f_\text{c,i}\right) -10 \log_\text{10} \left[\pi \Delta f^3\right].
\end{align}
Accordingly, the estimate of $f_\text{c,i}$ from $M$ data points in a section is given by
\begin{align}
    \hat{f}_\text{c,i} = 10^{ \frac{1}{2(M-1)} \sum_{m=1}^{M} \log_{10} \left( 10^{\mathcal{L}_\text{i}(\Delta f_m)/10} \pi \Delta f_{m}^3 \right) }.
    \label{eq:3db_cut_off}
\end{align}
Here,  $\mathcal{L}(\Delta f_m)$ is the measured \ac{PSD} at frequency offset $\Delta f_m$ inside a section that is evaluated.
\\
The estimator for the $\SI{3}{dB}$ cut-off frequency \eqref{eq:3db_cut_off} is used on the sections 1. and 3. where the reference oscillator and \ac{VCO} are dominant.
For the measurement displayed in Fig.~\ref{fig:pn_lin_regr}, this yields the estimate for the $\SI{3}{dB}$ cut-off frequency of reference oscillator and \ac{VCO}
\begin{align*}
    \hat{f}_\text{c,REF} = \SI{0.58}{Hz},~
    \hat{f}_\text{c,VCO} = \SI{630}{Hz}.
\end{align*}
As the oscillator frequency $f_0$ during the measurement is known and cut-off frequency estimate is established, the oscillator constant can be derived by $\hat{c} = \frac{\hat{f}_\text{c}}{\pi f_0^2}$, 
resulting in 
\begin{align*}
    \hat{c}_\text{REF} = \SI{4.58e-20}{s},~
    \hat{c}_\text{VCO} = \SI{5.01e-17}{s}.
\end{align*}
\begin{figure}[tb]
    \centering
    \includegraphics[width=\linewidth]{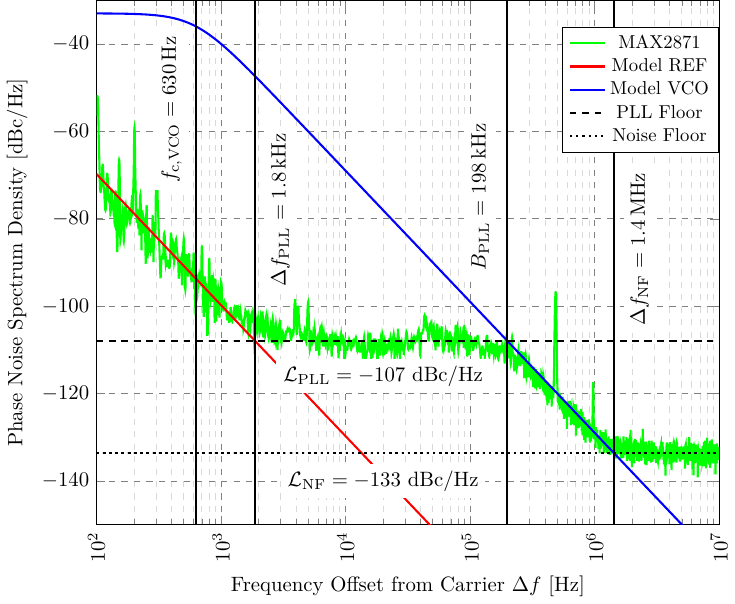}
    \caption{Measured phase noise spectrum of \ac{USRP} 2944R with parameter estimates and fitted models for reference oscillator and \ac{VCO} \eqref{eq:vco_model}}
    \label{fig:pn_param_est}
\end{figure}

The next objective is to estimate the in-band cut-off frequency $\Delta f_\text{PLL}$, \ac{PLL} bandwidth $B_\text{PLL}$, and the noise floor cut-off frequency $f_\text{NF}$.
To achieve this, the power level of the \ac{PLL} in-band noise and the noise floor is estimated.
This power level estimate is obtained via the sample mean estimator
\begin{align}
    \hat{\mathcal{L}} = \frac{1}{M-1} \sum_{m=1}^{M} \mathcal{L}_{m}.
\end{align}
Using this estimator, the estimated power levels for transition interval $\hat{\mathcal{L}}_\text{TR}$ and noise floor $\hat{\mathcal{L}}_\text{NF}$ are calculated from the corresponding sections 2. and 4. of the \ac{PSD}
\begin{align*}
    \hat{\mathcal{L}}_\text{PLL} = -107.9~\text{dBc/Hz},~
    \hat{\mathcal{L}}_\text{NF} = -133.7~\text{dBc/Hz}.
\end{align*}
The desired frequencies can then be found at the intersection point of the estimated power levels $\hat{\mathcal{L}}$ and the \ac{LP} filters modeling the reference oscillator and \ac{VCO} \eqref{eq:vco_model}.
The terms in \eqref{eq:vco_model} are rearranged to isolate the offset frequency $\Delta f$.
Then the estimator for the frequency can be written as
\begin{align}
    \Delta\hat{f} = \hat{f}_\text{c} \sqrt[3]{\left(\frac{1}{10^{\hat{\mathcal{L}}/10} \pi \hat{f}_\text{c}} - 1 \right)}.
\end{align}
By inserting $\hat{f}_\text{c} = \hat{f}_\text{c,REF}$ and $\hat{\mathcal{L}} = \hat{\mathcal{L}}_\text{PLL}$, the in-band cut-off frequency $\Delta\hat{f}_\text{PLL}$ is found
\begin{align*}
    \Delta\hat{f}_\text{PLL} = \SI{1865.7}{Hz}.
\end{align*}
In similar fashion, the \ac{PLL} bandwidth $\hat{B}_\text{PLL}$ and noise floor cut-off frequency $\Delta\hat{f}_\text{NF}$ can be calculated
\begin{align*}
    \hat{B}_\text{PLL} = \SI{197.9}{kHz},~
    \Delta\hat{f}_\text{NF} = \SI{1439.8}{kHz}.
\end{align*}

\begin{figure}[tb]
    \centering
    \includegraphics[width=\linewidth]{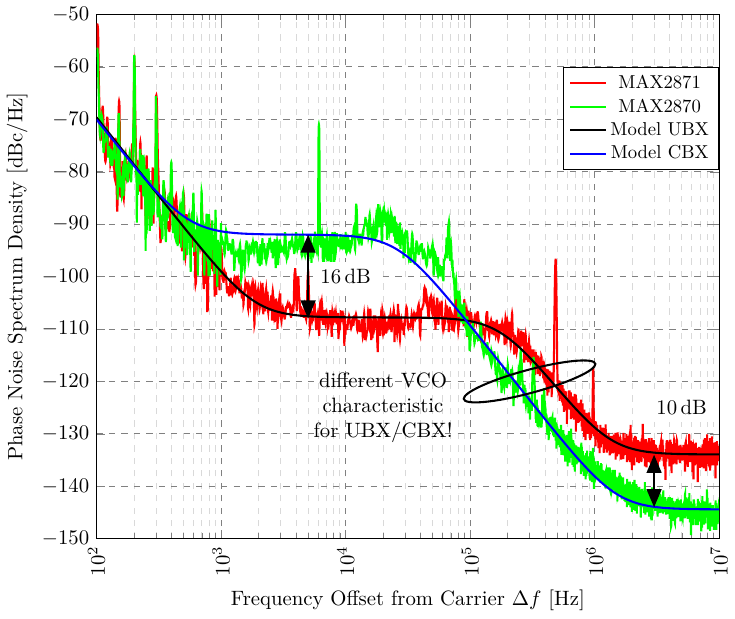}
    \caption{Measured phase noise spectra and fitted full model \eqref{eq:pn_model_simple} using the estimated parameters from Table \ref{table:parameter_estimates} for both daughterboard types.}
    \label{fig:pn_param_est_model}
\end{figure}

\begin{table}[ht]
    \centering
\begin{center}
        \resizebox{\linewidth}{!}{
\begin{tabular}{| l | l | l | l |}
 \hline
 Parameter & CBX / MAX2870 PLL & Parameter & UBX / MAX2871 PLL\\
 \hline
 $\hat{f}_\text{c,REF}$ & $0.5570 \pm 0.0249 ~ \text{Hz}$ & $\hat{f}_\text{c,REF}$ & $ 0.5853 \pm 0.0503 ~ \text{Hz}$ \\ 
 $\hat{c}_\text{REF}$ & $ (4.432 \pm 0.1978) \cdot10^{-20} ~ \text{s}$ &$\hat{c}_\text{REF}$ & $ (4.658 \pm 0.4005)\cdot10^{-20} ~ \text{s}$ \\
 $\hat{f}_\text{c,VCO}$ & $193.4 \pm 16.69 ~ \text{Hz}$ & $\hat{f}_\text{c,VCO}$ & $537.6 \pm 63.26 ~ \text{Hz}$ \\
 $\hat{c}_\text{VCO}$ & $ (1.539 \pm 0.1328)\cdot10^{-17} ~ \text{s}$ & $\hat{c}_\text{VCO}$ & $ (4.278 \pm 0.5034)\cdot10^{-17} ~ \text{s}$\\
 $\hat{\mathcal{L}}_\text{PLL}$ & $-91.9 \pm 1.735 ~ \text{dBc/Hz}$ & $\hat{\mathcal{L}}_\text{PLL}$  & $-107.8 \pm 0.7843 ~ \text{dBc/Hz}$\\
 $\Delta\hat{f}_\text{PLL}$ & $538.7 \pm 64.25~ \text{Hz}$ & $\Delta\hat{f}_\text{PLL}$ & $1872.1 \pm 119.61~ \text{Hz}$ \\
 $\hat{B}_\text{PLL}$ & $26.6 \pm 3.8 ~ \text{kHz}$ & $\hat{B}_\text{PLL}$ & $177.3 \pm 20.04~ \text{kHz}$\\
 $\hat{\mathcal{L}}_\text{NF}$ & $-144.4 \pm 0.2197 ~ \text{dBc/Hz}$ & $\hat{\mathcal{L}}_\text{NF}$ & $-134.0 \pm 0.1847~ \text{dBc/Hz}$\\
 $\Delta\hat{f}_\text{NF}$ & $1487 \pm 92.89~ \text{kHz}$ & $\Delta\hat{f}_\text{NF}$ & $1319 \pm 106.20~ \text{kHz}$\\
 \hline
\end{tabular}
}
\end{center}
\caption{Parameter estimates for different daughterboard types given as $\mu \pm \sigma$}
\label{table:parameter_estimates}
\end{table}

\subsection{Comparing Parameter Estimates for Different PLLs}
The procedure for estimating the phase noise model parameters is repeated for all \ac{USRP}s in the measured dataset \cite{data_bpmg_pc85_24}.
Parameter estimates are calculated as the average across all \ac{USRP}s of the same daughterboard type.

As expected, the parameter estimates displayed in \tabref{table:parameter_estimates} relating to the reference oscillator are identical for both daughterboards, since the same external reference is used for both measurements.
The measured \ac{VCO} oscillator constant differs between the two devices: $\hat{c}_\text{VCO} = \SI{1.539e-17}{s}$ for the MAX2870 (CBX) compared to $\SI{4.278e-17}{s}$ for the MAX2871 (UBX).
This is notable because the manufacturer datasheets report nearly identical VCO phase noise performance for both devices, with differences typically less than $\SI{2}{dB}$.
This finding reinforces a central theme of this work: datasheet specifications provide useful guidance, but measurement-based parameter extraction is essential for obtaining realistic models that reflect actual system-level performance.
A significant difference is observed in the in-band noise level: $\SI{-91.9}{dBc/Hz}$ for the CBX compared to $\SI{-107.8}{dBc/Hz}$ for the UBX.
The corresponding in-band corner frequencies also differ substantially: $\SI{538.7}{Hz}$ versus $\SI{1872.1}{Hz}$.
This is explained by superior in-band performance of the MAX2871 as stated by its datasheet which states normalized 1/f noise of $\SI{-122}{dBc/Hz}$ and in-band phase noise of $\SI{-102}{dBc/Hz}$, compared to the MAX2870 $\SI{-116}{dBc/Hz}$ and $\SI{-95}{dBc/Hz}$.
The \ac{PLL} bandwidth for the UBX daughterboard ($\SI{177.3}{kHz}$) is significantly larger than that of the CBX ($\SI{26.6}{kHz}$).
This is explained by the higher maximum \ac{PFD} frequency of $\SI{140}{MHz}$ for the MAX2871 (versus $\SI{105}{MHz}$ for the MAX2870) which enables wider loop bandwidths while maintaining lower in-band noise.
The measured noise floor shows a substantial difference between daughterboard models ($\SI{-144.4}{dBc/Hz}$ compared to $\SI{-134.0}{dBc/Hz}$).
This may reflect different output buffer designs or power amplifier stages on the daughterboards.

Figure \ref{fig:pn_param_est_model} shows that the fitted models accurately capture these characteristics, confirming that the extracted parameters provide a accurate representation of each \ac{PLL}'s phase noise behavior.

\section{Conclusion}\label{sec:conclusion}
%

This paper presents a comprehensive method for extracting phase noise model parameters from single-sideband \ac{PSD} measurements using the following approach: 1) measure the \ac{SSB} phase noise \ac{PSD}; 2) calculate the \ac{PSD} slope for \ac{PSD} segments using linear regression; 3) group segments into characteristic sections of the \ac{PLL}s \ac{PSD} based on their slope; 4) estimate phase noise model parameters with their respective \ac{LS} estimators; 5) obtain the fitted model by substituting parameter estimates, which can then be used for system-level simulations.
The model considered is based on an \ac{OU} process representation of the \ac{PLL} output, corresponds to a pole-zero spectrum in the frequency domain and is widely used in both time- and frequency-domain simulations.
The method was demonstrated on MAX2870 and MAX2871 \ac{PLL} synthesizers commonly found in SDR platforms such as the \ac{USRP} X310.
The extracted parameters accurately reproduce the measured \ac{PSD}, validating the model.

The parameter estimation method is general and can be readily applied to other \ac{PLL} frequency synthesizers, enabling system designers to obtain hardware-validated models for their specific devices.
Future works could exploit the modeling approach to compare simulation with measurement based performance assessment of \ac{MIMO} and JC\&S systems.




	\bibliographystyle{IEEEtran}
	\bibliography{references}
\end{document}